
\input jnl/jnl.tex
\input jnl/reforder.tex
\input jnl/eqnorder.tex
\title Transmission Resonance in an Infinite Strip of Phason-Defects
of a Penrose Approximant Network
\author  K. Moulopoulos and S. Roche
\affil  Laboratoire d' Etudes des Propri\'et\'es Electronique
des Solides, CNRS,  38042 Grenoble
\abstract
An exact method that  analytically  provides
 transfer  matrices
in finite  networks of quasicrystalline approximants of any dimensionality
is discussed.
We use  these matrices in two ways: a)
to exactly determine  the band structure of an
infinite approximant network
in analytical form;  b) to determine,
also analytically,  the
quantum resistance of a finite strip of a network under appropriate
boundary conditions.
As a result of a  subtle  interplay
between topology and phase interferences, we find that
a strip
of phason-defects
 along
a special symmetry direction
 of a low 2-d  Penrose approximant,
 leads to the rigorous vanishing of the
reflection coefficient
for certain energies. A similar behavior
appears in a low 3-d approximant.
This type of ``resonance" is discussed in
connection with the gap structure
of the corresponding ordered (undefected) system.

$$ $$

PACS numbers: 72.10.-d, 71.10.+x, 61.44.+p

\endtopmatter

\head{1. Introduction}

Recent experimental work\refto{aussois} shows that  quasicrystals
and quasicrystalline approximants have curious
(for metallic materials)
 transport
properties.
Anomalously high values of the low-temperature resistivity,
and the resistivity decrease with introduction of defects
or with increase of temperature, are typical examples.
We have recently analyzed\refto{roche} the role of a particular type of
phason-defect
on the conductance of a Fibonacci chain. From a consideration of the
scattering problem of noninteracting electrons through the chain
we have demonstrated subtle
interference effects between the hyperspace construction and the
phase-coherence of the wavefunctions in real space.
A full continuous formulation  (as
opposed to usual discetized approximations of a tight-binding form) was
critical\refto{roche} for obtaining
these interferences, i.e. the full phase-coherence
had to be  kept without truncation approximations.
Motivated by that work, we here follow a similar continuous scattering
formalism,
but  in model systems of higher dimensionality.

Within a continuous approach, a way to study $\it{exactly}$
structural aspects in higher dimensionality, is to treat $\it{networks}$ of
 wires\refto{avishai}
 that form a particular 2-d or 3-d structure. The electrons can propagate
only along the wires, but their wavefunctions $\it{split}$ at every vertex
by respecting continuity and quantum-mechanical current conservation
at every vertex.
(This splitting is the analog of scattering in the full high-d problem).
These are good (exactly soluble) model systems that can show non-trivial
interference effects on the phase of the exact  wavefunctions,
and that can also demonstrate how these are affected by the particular topology
(manifested by the connectivities characterizing the selected structure).

Our goal here is to study such structural effects on phase interferences,
but also
to see $\it{how}$ these are influenced by introduction
of $\it{defects}$ in the structure, in the form of disruption of long-range
order.
We find a type of resonance upon introduction of a
particular kind of defect, which is the result of a subtle interplay between
phase interferences
and the topology introduced by the defect.

The structures used can be periodic, but with non-trivial unit cells.
Periodicity combined with the scattering formalism leads to a natural
 study of quasicrystalline approximants.
We should emphasize that consideration
 of such a scattering (or equivalently, transmission) problem,
 gives $\it{full}$ information  on the  states and the  spectrum.
 The band structure of these
approximants can be given exactly in closed form (with the coupling of modes in
different directions kept in full without approximations).
These are therefore good textbook examples of derivations of
band structures in high dimensionality, by simply  starting from a scattering
problem
through a single unit cell.

The defects introduced are of a phason-type (but different from the ones
studied in ref.2) and in real space they
correspond to  $\it{flips}$ of the internal structure of the unit
cell. Our goal is to study the influence of such a flip on the transmission
properties. We actually find, as already mentioned,
 at least one system where the flipping
causes a rigorous vanishing of backscattering (reflection coefficient)
for some incidence energies.
Introduction of an infinite strip of such flippings leads to a ``resonance",
while introduction of a finite number of them lowers the resistance of an
otherwise ordered strip.
Although these results are rigorous for a strip of unit cells under some
special type
of boundary conditions, comparison with the same method but  applied to a
square lattice
gives evidence that this vanishing may be related to creation of some extended
states in the infinite lattice upon introduction of flipping. This observation
may be seen as a manifestation of anomalous transmission behavior,
consistent with numerical works\refto{fujiwara} on 2-d Penrose lattices (in a
tight-binding approximation),
and also in general agreement with  experiments\refto{aussois} in real
approximants.
 We also find that the special energies where the resonance occurs
lie in the middle of gaps of the infinite (undefected) approximant.
We propose this as an evidence that a possible creation of extended states may
be due to
the creation of levels (upon flipping) in  gaps of the ordered system.

\head{2. The method}

We solve the full  continuous and static  Schr\"odinger equation in each branch
(wire) of the network. No truncation approximations are made, so that
the full phase-coherence of the wavefunctions is maintained.
For simplicity, we use a vanishing potential along each wire (free electrons
along each branch) and take the length of each branch as common ($l$) (however
we
can still form various non-trivial structures
by connecting the wires appropriately). For more complicated treatments
(i.e. for effects of disorder by introducing random potentials along the
wires etc.) the above
assumptions  can of course be relaxed.

We then match the solutions from all wires that join at a point (vertex):
continuity of the wavefunction, and quantum-mechanical current conservation
at each vertex are enforced\refto{avishai}.
The method is formally similar to discussions\refto{degennes} of the linearized
Ginzburg-Landau equation in superconducting networks.

We give for example the matching equations for the
simplest single-vertex arrangements shown in Fig.1:

For case ($a$) of three wires we have
$$A_1 e^{i k l} +B_1 e^{-i k l}=A_1^\prime+B_1^\prime=A_0+B_0 \eqno(cont)$$
from  the continuity of the wavefunctions, and
$$A_1 e^{i k l}-B_1 e^{-i k l}=A_1^\prime-B_1^\prime+A_0-B_0 \eqno(current)$$
as one way to assure quantum-mechanical current conservation. In the above, $A$
and $B$ are the coefficients of
the two linearly independent plane waves (with wavevector $k$) along each wire.

A $2 \times 2$ transfer matrix can be defined that describes the scattering of
electrons incident at branch $1$ and transmitted at branch $1^\prime$, while
branch $0$ is viewed as ``transverse" escape of electrons. If for example
we look at a situation where electrons are $\it{not}$ incident from outside
through branch $0$
(which is equivalent to setting $B_0 =0$), but they can $\it{only}$ escape
through this branch
(i.e. $A_0 \neq 0$), then elimination of $A_0$ from equations \(cont) and
\(current) leads to
$$ \pmatrix {
A_1^\prime \cr
B_1^\prime \cr }= {1 \over 2} \;
\pmatrix{
e^{i k l} & -e^{-i k l} \cr
e^{i k l} & 3 e^{-i k l} \cr}.
\pmatrix{
A_1 \cr
B_1 \cr}  \eqno(transfer3)$$
which defines the $2 \times 2$ transfer matrix for this simplest possible case
(the determinant of this matrix is $1$, as expected, from conservation of
probability).

It is important to emphasize that the topology of the selected structure
(manifested by the corresponding connectivity of the wires) is crucial
for the {\it form} of the resulting matrices. To show this, let us also
determine
the $2 \times 2$ transfer matrix associated with case ($b$) with two transverse
channels  of Fig.1. For this
case, the matching equations are
$$A_1 e^{i k l} +B_1 e^{-i k l}=A_1^\prime+B_1^\prime=A_0+B_0=A_0^\prime+
B_0^\prime  \eqno(cont4)$$
from  the continuity of the wavefunctions, and
$$A_1 e^{i k l}-B_1 e^{-i k
l}=A_1^\prime-B_1^\prime+A_0-B_0+A_0^\prime-B_0^\prime  \eqno(current4)$$
from  the current conservation. Once again, if we treat branch $1$ as the input
branch, $1^\prime$ as the output branch, and $0$ and $0^\prime$ as free
transverse escapes (i.e. we impose $B_0 = B_0^\prime =0$, i.e. no additional
input of electrons from the transverse directions, but $\it{only}$ output that
is
naturally determined from the Schr\"odinger equation), then elimination of
$A_0$ and $A_0^\prime$ from equations
\(cont4) and \(current4) leads to
$$ \pmatrix {
A_1^\prime \cr
B_1^\prime \cr }=
\pmatrix{
0 & -e^{-i k l} \cr
e^{i k l} & 2 e^{-i k l} \cr}.
\pmatrix{
A_1 \cr
B_1 \cr}.  \eqno(transfer4)$$
Note that the $2 \times 2$ transfer matrix for this case is different from
\(transfer3),
the sole reason being the fact that
 there is one $\it{additional}$ transverse channel compared to the case ($a$).
(However the determinant of \(transfer4) is also $1$, as expected).

We conclude from these trivial (single-vertex) cases, that  the topology
(connectivity) is $\it{very}$
 important for the form of the transfer matrices (and hence of any transmission
properties, as, for example, the resistance (see below)). The selected
 topology actually leads
to nontrivial forms for the transfer matrices
and the  transmission properties, as will be discussed below.

It is important to note that even for a complex problem (of many wires
and many vertices
forming a particular  structure) it is always legitimate to
choose one wire as the input channel and  any other wire
as the output one, with all the rest  ``outside" wires (i.e. lying in the
exterior of the chosen unit)
treated as transverse escape-channels, irrespective of their number.
By then choosing boundary conditions of some type
for the coefficients in  these transverse wires,
we can always obtain a $2 \times 2$ transfer matrix, as above, but, in general,
with
elements that can be  quite complicated functions
  of $k l$,  the form of which depends on the
topology of the
structure (of $\it{both}$ internal $\it{and}$ external connectivities)
and describes the complete phase-coherence throughout the
selected unit.
This will be done in the following sections.

Let us first however use the concept of the transfer matrix, as introduced
above,
 to discuss transmission properties.
Usually, for a complex system with many channels,
one speaks in terms of the
multi-channel Landauer formula\refto{multi} in order to define a measure of the
resistance.
Here however we find it more convenient to focus on the channels that we
choose as input and output branches, and to define an $\it{effective}$
resistance,
even in the
 presence of the transverse escapes. By making a small
modification\refto{costas} to the
standard Landauer argument\refto{landauer} we find that this effective
resistance is given, for spin-${1 \over 2}$ electrons, by
$${\cal R} = {{\pi \hbar}\over e^2} \;\; {{(R-T+1)}\over{2 (1-R)}}
\eqno(resistance)$$
with $R$ and $T$ the (partial) reflection and transmission coefficients
resulting
from the $2 \times 2$ transfer matrix (as discussed above) of the problem under
consideration. It turns out that these
coefficients are given by the elements of the transfer matrix ($\hat{T}$)
 as follows:
$$R={{\mid T_{21} \mid ^2}\over
{\mid T_{22} \mid ^2 }} \eqno(R)$$
and
$$ T= {1 \over
{\mid T_{22} \mid ^2 }}. \eqno(T)$$

Note that \(resistance) yields the usual Landauer ratio $R/T$ if there are
no transverse escapes or if periodicity is enforced in the transverse channels
(or generally whenever $R+T=1$ is valid).

The  boundary conditions (for the transverse wires)  that we use in what
follows
are of two types: One might be called ``free scattering" and correspond to
vanishing transverse incidence (as the ones used
in the above trivial single-vertex  cases), and the other
is periodic boundary conditions in the transverse directions.
The latter can lead to band structures, as  will be  seen below.
But a link between the two types is expected, as we now discuss
in the case of the simplest 2-d structure: a square network.

\head{3. Example: Square Lattice}

\underbar{A. Transmission Problem}

The basic vertex problem to be solved is shown in Fig.1-d.
Matching the solution in all $N$ vertices results in
transfer matrices, with forms depending on the boundary
conditions.

1) For ``free scattering", namely $C_0=D_N=0$ (no current input transversely,
as discussed in previous Section) we obtain the following $2 N \times 2 N$
transfer matrix (for $N \geq 2$)
$$\hat{T}={1 \over {2 i \sin{k l}}}
\pmatrix{
e^{2 i k l}\;\; \hat{1}_N +e^{2 i k l}\;\; \hat{\Omega}    &
e^{-2 i k l}\;\; \hat{1}_N + \hat{\Omega} \cr
- \hat{1}_N - e^{2 i k l}\;\; \hat{\Omega} &
(1-2 e^{-2 i k l})\;\; \hat{1}_N - \hat{\Omega} \cr}  \eqno(freesquare)$$
with $\hat{1}_N$ denoting the $N \times N$ identity matrix, and
with $\hat{\Omega}$ being an $N \times N$ submatrix of the form
$$\hat{\Omega}=
\pmatrix{
0&-e^{-i k l}&0&0&0&...&0\cr
-e^{-i k l}&1&-e^{-i k l}&0&0&...&0\cr
0&-e^{-i k l}&1&-e^{-i k l}&0&...&0\cr
0&.&.&.&.&.&0\cr
0&.&.&.&.&.&0\cr
0&0&0&.&-e^{-i k l}&1&-e^{-i k l} \cr
0&0&0&.&0&-e^{-i k l}&0\cr
}.        \eqno(freeOmega)$$
Hence the total tranfer matrix of a finite piece of a square network
of ``width" $N$ and ``length" $M$
will be a product of $M$ matrices of the form \(freesquare).
This matrix summarizes the full information of multiple splittings
and it  can give all the transmission properties of this network
under this type of boundary conditions.
(We will see below, however, that a simple choice of width $N=2$ and
length $M=2$ (equivalent to a single unit cell)
is sufficient to exhibit resonance phenomena related to properties of the
infinite
square network).

2) For periodic boundary conditions along the transverse direction,
namely $C_N=C_0 \; e^{i N \phi}$ and $D_N=D_0 \; e^{i N \phi}$ we obtain the
 corresponding $2 N \times 2 N$ result (again for $N \geq 2$)
$$\hat{T} ={1 \over {2 i \sin{k l}}}
\pmatrix{
2 e^{2 i k l}\;\; \hat{1}_N -e^{i k l}\;\; \hat{\Omega}^\prime &
(1+e^{-2 i k l})\;\; \hat{1}_N - e^{- i k l}\;\; \hat{\Omega}^\prime \cr
-(1+e^{2 i k l})\;\; \hat{1}_N + e^{ i k l}\;\; \hat{\Omega}^\prime  &
-2 e^{-2 i k l}\;\; \hat{1}_N +e^{-i k l}\;\; \hat{\Omega}^\prime \cr
} \eqno(persquare)$$
with $\hat{\Omega}^\prime$ being again an $N \times N$ submatrix, but now  of
the form
$$\hat{\Omega}^\prime=
\pmatrix{
0&1&0&...&e^{-i \phi}\cr
1&0&1&...&0\cr
0&1&0&...&1\cr
.&.&.&.&.\cr
.&.&.&.&.\cr
e^{i \phi}&0&1&...&0\cr
}         \eqno(perOmega)$$
and once again the total tranfer matrix of a finite piece of a square network
($N \times M$) will be a product of $M$ matrices of the form \(persquare).

We make the following important observation for case 1):
if we take just a double vertex ($N=2$),  the transfer matrix
associated with the unit cell of the square network (hence $M=2$ as well)
is just the square of matrix \(freesquare). This matrix has off-diagonal
elements proportional to $\cot{k l}$, which are therefore vanishing
for $k l =(2 \rho+1) {\pi \over 2}$. These ``resonances" that are observed
for the $\it{single}$ unit cell under ``free-scattering" boundary conditions,
correspond, as we will show below, to a special property of the bands of
the $\it{infinite}$ square network and they are related  to  states
that are ``maximally extended" (see next Subsection).
Hence the finite (even small, with just a single unit cell)
transmission problem, for this type of boundary conditions,
 carries the memory of the most
extended states (in the thermodynamic limit)
in a form of vanishing of appropriate partial reflection coefficients.
This motivates the treatment of a single unit cell of a somewhat more
complex network in  Section 4.

\underbar{B. Band Structure}

Let us actually find the exact band structure of an (infinite) square network:
This results from relating coefficients  periodically in both
orthogonal directions.
By identifying, for example, in the single-vertex system
 ($N=M=1$) (see Fig.1-c)
the  coefficients in the following way
$$A_1^\prime = A_1 e^{i \phi_1},\;\;\;\;\;\;
B_1^\prime = B_1 e^{i \phi_1} $$
and
$$A_0^\prime = A_0 e^{i \phi_2},\;\;\;\;\;\;
B_0^\prime = B_0 e^{i \phi_2}, $$
all 8 coefficients appearing in this problem  can be eliminated, with a
resulting relation
between $k l$ and $\phi_1$ and $\phi_2$.
This relation is the exact band structure
(since $k$ is related to the energy $E$ by
$k^2={{2 m E}\over \hbar^2}$,  and the two phases define the crystal momenta
($q_1$ and $q_2$) in the two orthogonal directions through
$\phi_1=q_1 l$ and
$ \phi_2=q_2 l.$
The resulting band structure is
$${{2 m}\over \hbar^2}\; E\; l^2 =
\Biggl( \arccos{\bigl({1 \over 2} \cos{q_1 l}+{1 \over 2} \cos{q_2 l}
\bigr)} \Biggr)^2. \eqno(bandsquare)$$

Plotting \(bandsquare) for $q_1 l$ as a function of $E$ (for several fixed
values of $q_2 l$), we see gaps opening at regions where $q_1 l$ becomes a
complex
number. (We also observe a manifestation of the Higgs mechanism (i.e.
opening of a gap
at $E = 0$) that also appears and is further discussed in the next Section).
It is interesting that, by smoothly changing $q_2 l$ we see that the
band structure is actually $\it{moving}$.
This is an additional feature (of coupling between the two modes)
in comparison to standard tight-binding band structures.
Note, however, that during these movement, the special points $k l =(2 \rho+1)
{\pi \over 2}$
$\it{always}$ lie in bands (and they are the only points that have this
property) (see Fig.2). This shows that these points label what could be called
``maximally  extended" states.
Recall that these are the
points that give ``resonances"
(vanishings of the partial reflection coefficients) in the
``free-scattering" transfer
matrices of a $\it{single}$ unit cell.
This
 motivates our later proposition that
a resonance that we will find in  a non-trivial
network,  under introduction of a defect, may also be related to possible
creation of extended states.
(Spectral properties of a more general rectangular lattice has been recently
discussed\refto{exner}, with emphasis on incommensurability issues
with respect to the ratio of the two lattice constants).

\head{4. A
Low 2-d Penrose Approximant}

\underbar{A. Transmission Problem}

The  simplest  way of filling the plane periodically
 with the standard two Penrose rhombic tiles is shown in Fig.3-a
(where also the two linearly independent directions are shown, labelled
by the two phases $\phi_1$ and $\phi_2$ that will later enter in
the band structure).
This structure can be viewed as a low 2-d Penrose approximant.

Let us determine the transfer matrix for the scattering of electrons
through a single
unit cell of such an ordered system,
by treating branch $1$ as the input branch, and $1^\prime$ as the output one
(see Fig.3-a)). All other ``outside" branches shown in Fig.3-a are treated
as transverse escapes. Let us therefore  apply ``free-scattering"
boundary conditions for these escape-wires (that are $10$ in number).
It turns out that we now have overall $30$ matching equations (describing
the complete physics at the $7$ vertices).
The total number of branches is $21$ (which means $42$ coefficients overall),
and since from the boundary conditions $10$ of them vanish,
we can solve this linear system of equations for $(42-10)-30=2$
quantities (i.e. $A_1^\prime$ and $B_1^\prime$), a procedure  which indeed
yields a
$2 \times 2$ transfer matrix (linearly relating $A_1^\prime, B_1^\prime$
with $A_1,B_1$) as claimed earlier.
The above procedure is actually equivalent to inverting
  a $30 \times 30$
matrix, in order to obtain the final $2 \times 2$
transfer matrix for this problem. This $30 \times 30$ matrix carries the full
memory of
the structure of the unit cell (through the particular connectivities),
$\it{both}$ the internal structure, but $\it{also}$ the external connectivities
as well
(recall from earlier discussion in Section 2 that this was crucial for the form
of the resulting matrix).

The  transfer matrix that results from this procedure
 gives, through equations  \(R) and \(T), the following
exact results for the reflection and transmission coefficients
$$R= {1 \over 9}\;\;{
{31913+39744 \cos{2 k l}+8343 \cos{4 k l}}
\over
{5897+6336 \cos{2 k l}+567 \cos{4 k l}}} \eqno(2dR)$$
and
$$T= {128 \over 9}\;\;{
{13+12 \cos{2 k l} }
\over
{5897+6336 \cos{2 k l}+567 \cos{4 k l}}}. \eqno(2dT)$$
Note that $R$ has a maximum for
$k l= (2 \rho +1) {\pi \over 2}$.
This is actually related to the fact that for the infinite approximant, these
points
correspond to the middle of  gaps, as will be seen below (Subsection B).
Indeed for a strip of infinite such (ordered) units
parallel to the $1 1^\prime$-direction
 (i.e. in such a way as to have
 the output
of one as the input of another), the total transfer matrix is an infinite
product of elementary transfer matrices, and the corresponding $(12)$-element
 will diverge;
a   reflection
coefficient that goes to infinity indeed signifies the presence of a gap.

What is however more interesting is the problem of a unit with a
``flipped" internal structure (while keeping the external connectivities
$\it{the}$ $\it{same}$, to simulate the fact that the flipping is in the
interior of
$\it{only}$ one unit cell and not in the external environment). This ``flipped"
system is shown in Fig.3-b. By going through the new matching equations
and inverting the new $30 \times 30$ matrix we obtain the following
exact  results
$$R=
{{4 ({\cos{k l}})^2} \over 9}\;\;
{
{12781+8299 \cos{2 k l}-180 \cos{4 k l}-900 \cos{6 k l}}\over
{6821+6284 \cos{2 k l}+255 \cos{4 k l}-460 \cos{6 k l}-100 \cos{8 k l}}}
\eqno(flipR)$$
and
$$T={128 \over 9}\;\;{
{17+8 \cos{2 k l}}\over
{6821+6284 \cos{2 k l}+255 \cos{4 k l}-460 \cos{6 k l}-100 \cos{8 k l}}}.
\eqno(flipT)$$

{}From the structure of \(flipR) we see that for the special points
$k l =(2 \rho +1) {\pi \over 2}$ (where the ordered system was found to have
a maximum) the reflection coefficient $\it{vanishes}$
$\it{rigorously}$. This vanishing is a very special situation that does not
happen in the
ordered system. It shows that if we put any finite number of flipped units
in an otherwise ordered strip, the resistance will go down.
Even more spectacular is the fact that, even an infinite strip of
flipped units along this special symmetry direction
 will have a vanishing total reflection coefficient  (due to the
vanishing of the ($12$)-element of each transfer matrix, any product of
these matrices, even of infinite of them, will have a ($12$)-element
that will also be rigorously vanishing).
The reflection coefficients \(2dR) and \(flipR) are plotted in Fig.4.

\underbar{B. Band Structure}

Let us now determine the band structure of the ordered infinite approximant
network. We relate coefficients, through a phase $\phi_1$ or $\phi_2$ or
$\phi_1 - \phi_2$, for corresponding branches of unit cells
that repeat in the three possible directions.
An example is
 shown in Fig.5, showing connections    through phases
$\phi_1$ and $\phi_2$. The final result comes, after elimination of all
coefficients, from a $12 \times 12$
determinant, and is
$${2m \over \hbar^2}\; E\; l^2 =
\Biggl( {1 \over 2} \arccos \biggl(
{{(-3+4 \cos{\phi_1}+4 \cos{\phi_2}+4 \cos{(\phi_1 - \phi_2)}}
\over 9} \biggr)  \Biggr)^2. \eqno(band2d)$$

This exact and analytically given  band structure is plotted in Fig.6
(for fixed $\phi_2$).
We see again the gap structure opening at regions where $\phi_1$ becomes a
complex
quantity. Once again we observe gaps in the origin,
which is a manifestation of the Higgs mechanism: this is expected
for $\it{any}$ network;
because even in the long-wavelength limit the space is full
 of ``holes" and never homogeneous, resulting in the disappearance
of Goldstone modes and the opening of a gap at the origin.

A contour-plot $E(\phi_1, \phi_2)$ of \(band2d) (surfaces of constant energy)
is also given in Fig.7.
This shows the possible  ``Fermi surfaces",
if the incidence energy is identified, as usually done,  with the Fermi energy.
(For free particles in full 2-d space
we would of course have homocentric circles).

Note that the resonance  upon flipping found above,
is $\it{always}$ in the middle of gaps of the ordered system.
If the resonance is indeed a demonstration of extendedness of states
 (as it is in the case
of a  square network discussed earlier),
one interpretation would be that new levels may be created in some gaps of the
ordered system upon introduction of the  phason-defect.
This is in agreement with numerical findings for a 2-d Penrose
system\refto{fujiwara}
in a  tight-binding model.

We finally report that a  calculation for a much  bigger unit  (that consists
of a central unit cell together with all $6$
neighboring unit cells, see Fig.8) gives the result also plotted in Fig.8.
We note that a very deep minimum close to zero is also observed in the case
that we only flip the central unit cell, whereas a very pronounced maximum
appears in the ordered (unflipped) system.

\head{5. A 3-d Penrose approximant network}

We search for the above resonance phenomena
 in a more complex three-dimensional Penrose network, with a unit cell
consisting of 4 rhombohedra of 2 types, and
 shown in Fig.9-a.
It turns out that the problem is richer\refto{costas}
than the corresponding 2-d network, but in one case we
 indeed  observe the same type of resonance upon flipping of the
internal rhombohedra (see Fig.9-b), but again keeping the external
connectivities the same.
This result
is shown in Fig.9-c, where the reflection coefficients of
the ordered and the flipped system
are compared.

\head{6. Conclusion}

We have found a vanishing reflection coefficient for a strip of flipped
units in a 2-d and also in a 3-d network of wires
for special incidence energies. The first obvious
question is whether this type of resonance is
experimentally detectable. With recent  advances in
microfabrication techniques
extremely narrow wires can actually be manufactured.
(Experiments in quasiperiodic superconducting networks\refto{pannetier}
have already been performed in the past).
 Alternatively, one can try the analogous
 acoustic experiments\refto{acoustic} to observe the above phenomena.
What is required for these types of resonances to be observed, is
reflectionless reservoirs in the transverse channels, so that the
``free-scattering" boundary conditions are satisfied.

{}From a theoretical point of view and in relation to quasicrystalline
approximants, we have argued that the above resonances may be related to
possible creation of extended states, upon flipping, in the gaps of the
ordered system. If this is true, it  may be seen as a possible mechanism
of lowering the resistance with introduction of defects, a tendency that is
in general agreement with real experiments\refto{aussois} on approximants
and with numerical treatments\refto{fujiwara}.

Finally, from an academic point of view, the above model systems are rare
(if not unique) examples of exactly soluble band structures in high
dimensionality.
Furthermore, using these exact band structures as ``input"
one can go further to build a semiclassical dynamics\refto{ash} to study
time-dependent propagation in these networks.
Effects of disorder can also be studied through introduction of
random potentials as already mentioned. Finally, introduction of a
magnetic field is possible (and the problem is still  soluble).
This can address questions related to the Quantum Hall Effect\refto{hall}
in quasicrystalline approximants, a subject that is still  open
to investigation.

 One of us (K.M.) acknowledges support from the European Union through the
Human Capital and Mobility Program.

\references


\refis{aussois}
For a review, see C. Berger in
``Lectures On Quasicrystals", ed. F. Hippert and D. Gratias (Les Editions
de Physique Les Ulis, 1994)

\refis{roche}
K. Moulopoulos and S. Roche, to be published in {\it Int. J. Mod. Phys. B}.

\refis{avishai}
O. P. Exner and P. Seba, {\it Rep. Math. Phys.}, {\bf 27}, 7 (1989);
J. E. Avron and L. Sadun, {\it Phys. Rev. Lett.}, {\bf 62}, 3082 (1989);
Y. Avishai and J. M. Luck, {\it Phys. Rev. B}, {\bf 45}, 1074 (1992)

\refis{fujiwara}
T. Fujiwara, S. Yamamoto, and G. Trambly de Laissardiere,
{\it Phys. Rev. Lett.}, {\bf 71}
(1993) 4166.

\refis{multi}
D. C. Langreth and E. Abrahams, {\it Phys. Rev. B}, {\bf 24}, 2978 (1981)

\refis{costas}
K. Moulopoulos, unpublished.

\refis{landauer}
R. Landauer, {\it Philos. Mag.}, {\bf 21} (1970) 863.

\refis{acoustic}
S. He and J. D. Maynard, {\it Phys. Rev. Lett.} {\bf 62}, 1888 (1989)

\refis{ash}
N. W. Ashcroft and N. D. Mermin, ``Solid State Physics", Holt-Saunders
(1976), chapter 12

\refis{hall}
See M. Kohmoto, {\it J. Phys. Soc. Jap.} {\bf 62}, 4001 (1993)

\refis{degennes} P. G. de Gennes, {\it C. R. Acad. Sci. B} {\bf 292}, 279
(1981);
S. Alexander, {\it Phys. Rev. B} {\bf 27}, 1541 (1983)

\refis{pannetier}
B. Pannetier {\it et al.}, {\it Phys. Rev. Lett.} {\bf 53}, 1845 (1984);
A. Behrooz {\it et al.}, {\it Phys. Rev. Lett.} {\bf 57}, 368 (1986);
M. A. Itzler {\it et al.}, {\it Phys. Rev. B} {\bf 47}, 14165 (1993)

\refis{exner}
P. Exner, {\it Phys. Rev. Lett.} {\bf 74}, 3503 (1995)

\endreferences

\figurecaptions

FIG.1 (a) Simplest single-vertex arrangement, with branch $1$ being viewed as
input, $1^\prime$ as output, and $0$ as ``transverse" branch.
(b) Single vertex with two transverse branches $0$ and $0^\prime$.
(c) Elementary vertex for the square network problem.
(d) Multi-vertex arrangement for a finite square network.

FIG.2 Movement of the band structure of a square network as $q_2 l$ is
varied. Solid curve corresponds to $q_2 l=0$, Dashed-dotted  to $q_2 l=
{\pi \over 2}$, and
Dotted curve  to $q_2 l={\pi}$. The two vertical lines show the positions of
two
special energies corresponding to $k l={\pi \over 2}$ and $k l={{3 \pi}\over
2}$.
Note that these points  {\it always} belong to a band. (Units ${{2 m}\over
\hbar^2}\; l^2 =1$ are used).

FIG.3 (a) A low 2-d approximant network. There are three directions
along which the unit cell is repeated; two of them are independent
and are symbolized by the phases $\phi_1$ and $\phi_2$ (the third is then
described by the phase
$\phi_1-\phi_2$).
A single unit cell is shown in the center, with  arrows in the
``outside branches" (which are  determined by filling the plane with
 periodic repetition of this unit cell
in all three directions). Single arrows signify ``free-scattering"
boundary conditions (see text, beginning of Section 4).
$1$ and $1^\prime$ denote input and output branches (where reflections are
allowed).
(b) The result of flipping the interior of the central unit cell,
but keeping the external
environment the same.

FIG.4 Reflection coefficients, equ.'s \(2dR) and \(flipR), for the units
shown in Fig.3(a) and 3(b). Unit 3(a) shows  a local maximum, and unit 3(b)
shows a vanishing at the special point $k l={\pi \over 2}$.

FIG.5 Example of how to relate coefficients through $\phi_1$ and $\phi_2$,
in order to determine analytically the band structure of an infinite network
with the unit cell of Fig. 3-a.

FIG.6 Band structure of an infinite network with the unit cell of Fig. 3-a
(for $\phi_2 = {\pi \over 2}$).
Gaps open at regions where $\phi_1$ becomes a complex number.

FIG.7 Possible Fermi surfaces associated with the network with the unit cell
of Fig. 3-a.

FIG.8 Reflection coefficients for a bigger central unit (shown on top left),
consisting of $7$ unit cells (one central and the $6$ neighboring ones).
For the ordered  (unflipped) system (left) we observe a local maximum;
for the system with a flipped central unit cell (right), we observe a deep
minimum
very close to zero ($R_{min} =0.017$),
at $k l ={\pi \over 2}$.

FIG.9 (a) The unit cell of a low 3-d approximant.
(b) The corresponding unit with ``flipped" internal structure
(but the same external environment).
(c) The corresponding reflection coefficients showing a vanishing
for the flipped system at the same special point $k l ={\pi \over 2}$.

\endfigurecaptions

\end

\\
\catcode`@=11
\newcount\tagnumber\tagnumber=0

\immediate\newwrite\eqnfile
\newif\if@qnfile\@qnfilefalse
\def\write@qn#1{}
\def\writenew@qn#1{}
\def\w@rnwrite#1{\write@qn{#1}\message{#1}}
\def\@rrwrite#1{\write@qn{#1}\errmessage{#1}}

\def\taghead#1{\gdef\t@ghead{#1}\global\tagnumber=0}
\def\t@ghead{}

\expandafter\def\csname @qnnum-3\endcsname
  {{\t@ghead\advance\tagnumber by -3\relax\number\tagnumber}}
\expandafter\def\csname @qnnum-2\endcsname
  {{\t@ghead\advance\tagnumber by -2\relax\number\tagnumber}}
\expandafter\def\csname @qnnum-1\endcsname
  {{\t@ghead\advance\tagnumber by -1\relax\number\tagnumber}}
\expandafter\def\csname @qnnum0\endcsname
  {\t@ghead\number\tagnumber}
\expandafter\def\csname @qnnum+1\endcsname
  {{\t@ghead\advance\tagnumber by 1\relax\number\tagnumber}}
\expandafter\def\csname @qnnum+2\endcsname
  {{\t@ghead\advance\tagnumber by 2\relax\number\tagnumber}}
\expandafter\def\csname @qnnum+3\endcsname
  {{\t@ghead\advance\tagnumber by 3\relax\number\tagnumber}}

\def\equationfile{%
  \@qnfiletrue\immediate\openout\eqnfile=\jobname.eqn%
  \def\write@qn##1{\if@qnfile\immediate\write\eqnfile{##1}\fi}
  \def\writenew@qn##1{\if@qnfile\immediate\write\eqnfile
    {\noexpand\tag{##1} = (\t@ghead\number\tagnumber)}\fi}
}

\def\callall#1{\xdef#1##1{#1{\noexpand\call{##1}}}}
\def\call#1{\each@rg\callr@nge{#1}}

\def\each@rg#1#2{{\let\thecsname=#1\expandafter\first@rg#2,\end,}}
\def\first@rg#1,{\thecsname{#1}\apply@rg}
\def\apply@rg#1,{\ifx\end#1\let\next=\relax%
\else,\thecsname{#1}\let\next=\apply@rg\fi\next}

\def\callr@nge#1{\calldor@nge#1-\end-}
\def\callr@ngeat#1\end-{#1}
\def\calldor@nge#1-#2-{\ifx\end#2\@qneatspace#1 %
  \else\calll@@p{#1}{#2}\callr@ngeat\fi}
\def\calll@@p#1#2{\ifnum#1>#2{\@rrwrite{Equation range #1-#2\space is bad.}
\errhelp{If you call a series of equations by the notation M-N, then M and
N must be integers, and N must be greater than or equal to M.}}\else%
 {\count0=#1\count1=#2\advance\count1
by1\relax\expandafter\@qncall\the\count0,%
  \loop\advance\count0 by1\relax%
    \ifnum\count0<\count1,\expandafter\@qncall\the\count0,%
  \repeat}\fi}

\def\@qneatspace#1#2 {\@qncall#1#2,}
\def\@qncall#1,{\ifunc@lled{#1}{\def\next{#1}\ifx\next\empty\else
  \w@rnwrite{Equation number \noexpand\(>>#1<<) has not been defined yet.}
  >>#1<<\fi}\else\csname @qnnum#1\endcsname\fi}

\let\eqnono=\eqno
\def\eqno(#1){\tag#1}
\def\tag#1$${\eqnono(\displayt@g#1 )$$}

\def\aligntag#1\endaligntag
  $${\gdef\tag##1\\{&(##1 )\cr}\eqalignno{#1\\}$$
  \gdef\tag##1$${\eqnono(\displayt@g##1 )$$}}

\def\eqalignno#1{\displ@y \tabskip\centering
  \halign to\displaywidth{\hfil$\displaystyle{##}$\tabskip\z@skip
    &$\displaystyle{{}##}$\hfil\tabskip\centering
    &\llap{$\displayt@gpar##$}\tabskip\z@skip\crcr
    #1\crcr}}

\def\displayt@gpar(#1){(\displayt@g#1 )}

\def\displayt@g#1 {\rm\ifunc@lled{#1}\global\advance\tagnumber by1
        {\def\next{#1}\ifx\next\empty\else\expandafter
        \xdef\csname @qnnum#1\endcsname{\t@ghead\number\tagnumber}\fi}%
  \writenew@qn{#1}\t@ghead\number\tagnumber\else
        {\edef\next{\t@ghead\number\tagnumber}%
        \expandafter\ifx\csname @qnnum#1\endcsname\next\else
        \w@rnwrite{Equation \noexpand\tag{#1} is a duplicate number.}\fi}%
  \csname @qnnum#1\endcsname\fi}

\def\ifunc@lled#1{\expandafter\ifx\csname @qnnum#1\endcsname\relax}

\let\@qnend=\end\gdef\end{\if@qnfile
\immediate\write16{Equation numbers written on []\jobname.EQN.}\fi\@qnend}

\catcode`@=12
\\

\font\twelverm=amr10 scaled 1200    \font\twelvei=ammi10 scaled 1200
\font\twelvesy=amsy10 scaled 1200   \font\twelveex=amex10 scaled 1200
\font\twelvebf=ambx10 scaled 1200   \font\twelvesl=amsl10 scaled 1200
\font\twelvett=amtt10 scaled 1200   \font\twelveit=amti10 scaled 1200

\skewchar\twelvei='177   \skewchar\twelvesy='60


\def\twelvepoint{\normalbaselineskip=12.4pt plus 0.1pt minus 0.1pt
  \abovedisplayskip 12.4pt plus 3pt minus 9pt
  \belowdisplayskip 12.4pt plus 3pt minus 9pt
  \abovedisplayshortskip 0pt plus 3pt
  \belowdisplayshortskip 7.2pt plus 3pt minus 4pt
  \smallskipamount=3.6pt plus1.2pt minus1.2pt
  \medskipamount=7.2pt plus2.4pt minus2.4pt
  \bigskipamount=14.4pt plus4.8pt minus4.8pt
  \def\rm{\fam0\twelverm}          \def\it{\fam\itfam\twelveit}%
  \def\sl{\fam\slfam\twelvesl}     \def\bf{\fam\bffam\twelvebf}%
  \def\mit{\fam 1}                 \def\cal{\fam 2}%
  \def\tt{\twelvett}
  \textfont0=\twelverm   \scriptfont0=\tenrm   \scriptscriptfont0=\sevenrm
  \textfont1=\twelvei    \scriptfont1=\teni    \scriptscriptfont1=\seveni
  \textfont2=\twelvesy   \scriptfont2=\tensy   \scriptscriptfont2=\sevensy
  \textfont3=\twelveex   \scriptfont3=\twelveex  \scriptscriptfont3=\twelveex
  \textfont\itfam=\twelveit
  \textfont\slfam=\twelvesl
  \textfont\bffam=\twelvebf \scriptfont\bffam=\tenbf
  \scriptscriptfont\bffam=\sevenbf
  \normalbaselines\rm}



\def\beginlinemode{\endmode
  \begingroup\parskip=0pt \obeylines\def\\{\par}\def\endmode{\par\endgroup}}
\def\beginparmode{\endmode
  \begingroup \def\endmode{\par\endgroup}}
\let\endmode=\par
{\obeylines\gdef\
{}}
\def\singlespace{\baselineskip=\normalbaselineskip}

\def\oneandahalfspace{\baselineskip=\normalbaselineskip
  \multiply\baselineskip by 3 \divide\baselineskip by 2}
\def\doublespace{\baselineskip=\normalbaselineskip \multiply\baselineskip by 2}

\newcount\firstpageno
\firstpageno=2
\footline={\ifnum\pageno<\firstpageno{\hfil}\else{\hfil\twelverm\folio\hfil}\fi}
\def\toppageno{\global\footline={\hfil}\global\headline
  ={\ifnum\pageno<\firstpageno{\hfil}\else{\hfil\twelverm\folio\hfil}\fi}}
\let\rawfootnote=\footnote		
\def\footnote#1#2{{\rm\singlespace\parindent=0pt\parskip=0pt
  \rawfootnote{#1}{#2\hfill\vrule height 0pt depth 6pt width 0pt}}}
\def\raggedcenter{\leftskip=4em plus 12em \rightskip=\leftskip
  \parindent=0pt \parfillskip=0pt \spaceskip=.3333em \xspaceskip=.5em
  \pretolerance=9999 \tolerance=9999
  \hyphenpenalty=9999 \exhyphenpenalty=9999 }
\def\dateline{\rightline{\ifcase\month\or
  January\or February\or March\or April\or May\or June\or
  July\or August\or September\or October\or November\or December\fi
  \space\number\year}}
\def\today{\ifcase\month\or
  January\or February\or March\or April\or May\or June\or
  July\or August\or September\or October\or November\or December\fi
  \space\number\day, \number\year}
\def\received{\vskip 3pt plus 0.2fill
 \centerline{\sl (Received\space\ifcase\month\or
  January\or February\or March\or April\or May\or June\or
  July\or August\or September\or October\or November\or December\fi
  \qquad, \number\year)}}


\hsize=6.5truein
\hoffset=0truein
\vsize=8.9truein
\voffset=0truein
\parskip=\medskipamount
\def\\{\cr}
\twelvepoint		
\doublespace		
\overfullrule=0pt	




\def\title			
  {\null\vskip 3pt plus 0.2fill
   \beginlinemode \doublespace \raggedcenter \bf}

\def\author			
  {\vskip 3pt plus 0.2fill \beginlinemode
   \singlespace \raggedcenter}

\def\affil			
  {\vskip 3pt plus 0.1fill \beginlinemode
   \oneandahalfspace \raggedcenter \sl}

\def\abstract			
  {\vskip 3pt plus 0.3fill \beginparmode
   \doublespace ABSTRACT: }

\def\submit  			
	{\vskip 24pt \beginlinemode
	\noindent \rm Submitted to: \sl}

\def\endtopmatter		
  {\endpage			
   \body}

\def\body			
  {\beginparmode}		

\def\head#1{			
  \goodbreak\vskip 0.5truein	
  {\immediate\write16{#1}
   \raggedcenter \uppercase{#1}\par}
   \nobreak\vskip 0.25truein\nobreak}

\def\beneathrel#1\under#2{\mathrel{\mathop{#2}\limits_{#1}}}

\def\refto#1{$^{#1}$}		

\def\references			
  {\head{References}		
   \beginparmode
   \frenchspacing \parindent=0pt \leftskip=1truecm
   \parskip=8pt plus 3pt \everypar{\hangindent=\parindent}}

\gdef\refis#1{\item{#1.\ }}			

\gdef\journal#1, #2, #3, 1#4#5#6{		
    {\sl #1~}{\bf #2}, #3 (1#4#5#6)}		

\def\endreferences{\body}

\def\figurecaptions		
  {\endpage
   \beginparmode
   \head{Figure Captions}
}

\def\endfigurecaptions{\body}

\def\endpage			
  {\vfill\eject}

\def\endpaper			
  {\endmode\vfill\supereject}


\def\heading				
  {\vskip 0.5truein plus 0.1truein	
   \beginparmode \def\\{\par} \parskip=0pt \singlespace \raggedcenter}

\def\subheading				
  {\vskip 0.25truein plus 0.1truein	
   \beginlinemode \singlespace \parskip=0pt \def\\{\par}\raggedcenter}

\def\tag#1$${\eqno(#1)$$}

\def\align#1$${\eqalign{#1}$$}

\def\aligntag#1$${\gdef\tag##1\\{&(##1)\cr}\eqalignno{#1\\}$$
  \gdef\tag##1$${\eqno(##1)$$}}

\def\endaligntag{}

\def\overset#1\to#2{{\mathop{#2}^{#1}}}
\def\underset#1\to#2{{\mathop{#2}_{#1}}}


\def\ref#1{Ref.~#1}			
\def\Ref#1{Ref.~#1}			
\def\[#1]{[\cite{#1}]}
\def\cite#1{{#1}}
\def\(#1){(\call{#1})}
\def\call#1{{#1}}
\def\taghead#1{}
\def\frac#1#2{{#1 \over #2}}

\def\12{{1\over2}}

\def\sla{\raise.15ex\hbox{$/$}\kern-.57em}
\def\leaderfill{\leaders\hbox to 1em{\hss.\hss}\hfill}
\def\twiddle{\lower.9ex\rlap{$\kern-.1em\scriptstyle\sim$}}
\def\bigtwiddle{\lower1.ex\rlap{$\sim$}}
\def\gtwid{\mathrel{\raise.3ex\hbox{$>$\kern-.75em\lower1ex\hbox{$\sim$}}}}
\def\ltwid{\mathrel{\raise.3ex\hbox{$<$\kern-.75em\lower1ex\hbox{$\sim$}}}}
\def\square{\kern1pt\vbox{\hrule height 1.2pt\hbox{\vrule width 1.2pt\hskip 3pt
   \vbox{\vskip 6pt}\hskip 3pt\vrule width 0.6pt}\hrule height 0.6pt}\kern1pt}
\def\tdot#1{\mathord{\mathop{#1}\limits^{\kern2pt\ldots}}}

\def\pmb#1{\setbox0=\hbox{#1}%
  \kern-.025em\copy0\kern-\wd0
  \kern  .05em\copy0\kern-\wd0
  \kern-.025em\raise.0433em\box0 }

\\
\catcode`@=11
\newcount\r@fcount \r@fcount=0
\newcount\r@fcurr
\immediate\newwrite\reffile
\newif\ifr@ffile\r@ffilefalse
\def\w@rnwrite#1{\ifr@ffile\immediate\write\reffile{#1}\fi\message{#1}}

\def\writer@f#1>>{}
\def\referencefile{
  \r@ffiletrue\immediate\openout\reffile=\jobname.ref%
  \def\writer@f##1>>{\ifr@ffile\immediate\write\reffile%
    {\noexpand\refis{##1} = \csname r@fnum##1\endcsname = %
     \expandafter\expandafter\expandafter\strip@t\expandafter%
     \meaning\csname r@ftext\csname r@fnum##1\endcsname\endcsname}\fi}%
  \def\strip@t##1>>{}}

\def\citeall#1{\xdef#1##1{#1{\noexpand\cite{##1}}}}
\def\cite#1{\each@rg\citer@nge{#1}}	

\def\each@rg#1#2{{\let\thecsname=#1\expandafter\first@rg#2,\end,}}
\def\first@rg#1,{\thecsname{#1}\apply@rg}	
\def\apply@rg#1,{\ifx\end#1\let\next=\relax
\else,\thecsname{#1}\let\next=\apply@rg\fi\next}

\def\citer@nge#1{\citedor@nge#1-\end-}	
\def\citer@ngeat#1\end-{#1}
\def\citedor@nge#1-#2-{\ifx\end#2\r@featspace#1 
  \else\citel@@p{#1}{#2}\citer@ngeat\fi}	
\def\citel@@p#1#2{\ifnum#1>#2{\errmessage{Reference range #1-#2\space is bad.}
    \errhelp{If you cite a series of references by the notation M-N, then M and
    N must be integers, and N must be greater than or equal to M.}}\else%
 {\count0=#1\count1=#2\advance\count1
by1\relax\expandafter\r@fcite\the\count0,%
  \loop\advance\count0 by1\relax
    \ifnum\count0<\count1,\expandafter\r@fcite\the\count0,%
  \repeat}\fi}

\def\r@featspace#1#2 {\r@fcite#1#2,}	
\def\r@fcite#1,{\ifuncit@d{#1}		
    \expandafter\gdef\csname r@ftext\number\r@fcount\endcsname%
    {\message{Reference #1 to be supplied.}\writer@f#1>>#1 to be supplied.\par
     }\fi%
  \csname r@fnum#1\endcsname}

\def\ifuncit@d#1{\expandafter\ifx\csname r@fnum#1\endcsname\relax%
\global\advance\r@fcount by1%
\expandafter\xdef\csname r@fnum#1\endcsname{\number\r@fcount}}

\let\r@fis=\refis			
\def\refis#1#2#3\par{\ifuncit@d{#1}
    \w@rnwrite{Reference #1=\number\r@fcount\space is not cited up to now.}\fi%
  \expandafter\gdef\csname r@ftext\csname r@fnum#1\endcsname\endcsname%
  {\writer@f#1>>#2#3\par}}

\def\r@ferr{\endreferences\errmessage{I was expecting to see
\noexpand\endreferences before now;  I have inserted it here.}}
\let\r@ferences=\references
\def\references{\r@ferences\def\endmode{\r@ferr\par\endgroup}}

\let\endr@ferences=\endreferences
\def\endreferences{\r@fcurr=0
  {\loop\ifnum\r@fcurr<\r@fcount
    \advance\r@fcurr by 1\relax\expandafter\r@fis\expandafter{\number\r@fcurr}%
    \csname r@ftext\number\r@fcurr\endcsname%
  \repeat}\gdef\r@ferr{}\endr@ferences}


\let\r@fend=\endpaper\gdef\endpaper{\ifr@ffile
\immediate\write16{Cross References written on []\jobname.REF.}\fi\r@fend}

\catcode`@=12

\citeall\refto		
\citeall\ref		%
\citeall\Ref		%

\\